\documentclass[aps,prl,reprint,amsmath,amsfonts]{revtex4-2}
\usepackage[colorlinks=true,allcolors=blue]{hyperref}
\usepackage[T1]{fontenc}
\usepackage[english]{babel}
\usepackage{braket}
\usepackage{bm}
\usepackage{microtype}

\begin{document}

\title{Why quarkonium hybrid coupling to two \textit{\textbf{S}}-wave heavy-light mesons is not suppressed}
\author{R. Bruschini}
\email{bruschini.1@osu.edu}
\affiliation{Department of Physics, The Ohio State University, Columbus, OH 43210, USA}

\begin{abstract}
We examine the couplings of quarkonium hybrids to heavy-light meson pairs in the Born-Oppenheimer approximation for QCD. The lowest hybrid multiplets consist of bound states of the $\Pi_u$ and $\Sigma_u^-$ potentials. We find that the $\Sigma_u^-$ potential can couple to pairs of $S$-wave mesons through string breaking, while the $\Pi_u$ potential cannot. From this observation, we derive model-independent selection rules that contradict previous expectations that quarkonium hybrids are forbidden to decay into pairs of $S$-wave mesons. These Born-Oppenheimer selection rules are consistent with the partial decay widths of the lowest charmonium hybrid with exotic quantum numbers $J^{PC}=1^{-+}$ recently calculated using lattice QCD.
\end{abstract}

\maketitle

While most observed hadrons can be accommodated as conventional quark-model states composed of three quarks or a quark-antiquark pair, QCD allows for the existence of exotic hadrons whose constituents include additional quarks, antiquarks, or gluons. The simplest exotic hadrons that contain a heavy quark-antiquark ($Q\bar{Q}$) pair also include a gluonic excitation and are called quarkonium hybrid mesons.

It is widely believed that hybrid mesons are forbidden to decay into two $S$-wave mesons. This idea, which was first proposed by Tanimoto in 1982 \cite{Tan82}, is rooted in constituent gluon models and flux tube models \cite{Tan82,LeY84,Is85,Idd88,Ish91,Clo94,Pag96,Pag98,Kou05}. The origin of this expectation was traced back by Page in a 1997 paper entitled ``Why hybrid meson coupling to two $S$-wave mesons is suppressed'' \cite{Pag96}. He derived a selection rule under the assumption that the $Q\bar{Q}$ pair is nonrelativistic using solely the symmetries of the decay amplitude. Page's selection rule states that the decay of a $Q\bar{Q}$ (quarkonium or quarkonium hybrid) meson into a pair of $S$-wave heavy-light mesons is suppressed if the $CP$-parity of the decaying meson, the total spin $S_{Q\bar{Q}}$ of the $Q\bar{Q}$ pair, and the total spin $S_{q\bar{q}}$ of the light quark-antiquark ($q\bar{q}$) pair created in the decay process satisfy
\begin{equation}
CP = (-1)^{S_{Q\bar{Q}} + S_{q\bar{q}} + 1}.
\label{pageq}
\end{equation}
Since a quarkonium hybrid meson has $CP=(-1)^{S_{Q\bar{Q}}}$, as explained later, Eq.~\eqref{pageq} implies that quarkonium hybrid decays into two $S$-wave heavy-light mesons are suppressed if $S_{q\bar{q}}=1$, which is the assumption of most hybrid meson decay models. 

Albeit motivated by models only, the hypothesis that $S_{q\bar{q}}=1$ in hybrid decays has somehow risen to the stage of conventional wisdom, and nowadays it is widely believed that quarkonium hybrid mesons are forbidden to decay into two $S$-wave mesons; see, for instance, Refs.~\cite{Mey15,Far20} and references therein. This belief seemed to have been validated by lattice QCD calculations of decays of the lightest hybrid meson with exotic quantum numbers $J^{PC}=1^{-+}$ in the cases of two light quark flavors \cite{Mcn06} and three light quark flavors with $SU(3)$ symmetry \cite{Wos20}. Its decay widths into two $S$-wave mesons are significantly smaller than its decay widths into a $P$-wave meson and an $S$-wave meson. However, a recent calculation of the decays of the lowest $1^{-+}$ charmonium hybrid using lattice QCD refutes this conventional wisdom \cite{Shi23}. The widths for its decays into the pairs of $S$-wave charm mesons $D^\ast\bar{D}$ and $D^\ast\bar{D}^\ast$ are smaller than but comparable to its width into the $P$-wave charm meson $D_1$ and the $S$-wave charm meson $\bar{D}$. This surprising difference between light hybrids and quarkonium hybrids calls out for a theoretical explanation.

In this Letter, we derive selection rules for transitions of quarkonium hybrids into pairs of $S$-wave heavy-light mesons in the Born-Oppenheimer (B\nobreakdash-O) approximation for QCD. These model-independent selection rules reveal that decays into two $S$-wave heavy-light mesons are allowed for many but not all quarkonium hybrid $J^{PC}$ quantum numbers. In particular, they are allowed for $J^{PC}=1^{-+}$.

The B\nobreakdash-O approximation for QCD provides a simple physical picture of quarkonium hybrids \cite{Jug99,Bal01}. A quarkonium or quarkonium hybrid meson is composed of a $Q\bar{Q}$ pair interacting with the light QCD fields for gluons and light quarks and antiquarks. Given that the masses of the $Q$ and $\bar{Q}$ are much larger than the nonperturbative QCD energy scale $\Lambda_\textup{QCD}$, the light QCD fields respond adiabatically to the motion of the $Q$ and $\bar{Q}$. The $Q$ and $\bar{Q}$ therefore act as static color sources with separation $\bm{r}$. The energy levels of light QCD with static sources can be calculated numerically using lattice QCD. They provide potentials that determine the $Q\bar{Q}$ meson spectrum through a (multichannel) Schr\"odinger equation. Rigorous effective field theories following the B\nobreakdash-O philosophy have been developed in recent years \cite{Ber15} and successfully applied to the spectrum of quarkonium hybrids \cite{Ber15,On17,Bra19,Sot23} and to their inclusive decay rates into quarkonium \cite{On17,Bra22}.

The static sources break the $SO(3)$ symmetry of QCD down to a cylindrical symmetry. The traditional B\nobreakdash-O quantum numbers of the light QCD fields in the presence of two static sources are $\Lambda_\eta^\epsilon$, with $\Lambda=\lvert\bm{\bm{J}_\textup{light}}\cdot\bm{\hat{r}}\rvert$ the modulus of the projection of the light QCD angular momentum $\bm{J}_\textup{light}$ onto the $Q\bar{Q}$ axis $\bm{\hat{r}}$, $\eta=g$ or $u$ for $(CP)_\textup{light}=+$ or $-$, and $\epsilon=\pm$ for the symmetry under reflections through a plane containing $Q\bar{Q}$ in the case $\Lambda=0$. The values of $\Lambda=0,1,2,\dots$ are conventionally labeled with capital Greek letters, $\Sigma,\Pi,\Delta$, and so on.

In pure $\mathrm{SU(3)}$ gauge theory, the ground-state static energy level with B\nobreakdash-O quantum numbers $\Sigma_g^+$ is identified with the conventional quarkonium potential. It approaches an attractive Coulomb potential at short distances and it increases linearly at long distances, similarly to the phenomenological Cornell potential for quarkonium \cite{Eic80}. Many of the excited-state static energy levels, associated to quarkonium hybrid potentials, were first calculated in pure $\mathrm{SU(3)}$ gauge theory in Ref.~\cite{Jug99}. The two lowest hybrid B\nobreakdash-O potentials, with B\nobreakdash-O quantum numbers $\Pi_u$ and $\Sigma_u^-$, are degenerate at short $Q\bar{Q}$ distances, where they behave like a repulsive Coulomb potential, then decrease to different minima at intermediate distances, and finally increase linearly at long distances. There have been several more recent calculations of the hybrid potentials in pure $\mathrm{SU(3)}$ gauge theory \cite{Cap18,Mul19,Sch21,Sha23}. The $\Pi_u$ potential below the string-breaking radius has also been calculated using lattice QCD with two flavors of light quarks in Ref.~\cite{Bal00}.

In QCD with dynamical light quarks, the ground-state $\Sigma_g^+$ static energy level approaches an attractive Coulomb potential at short distances but at long distances it approaches a constant equal to twice the energy of a static meson. This change in the behavior at long distance is referred to as string breaking, because it can be attributed to the breaking of a color flux tube connecting the $Q$ and $\bar{Q}$ by the creation of a light $q\bar{q}$ pair. The effects of string breaking on the lowest $\Sigma_g^+$ potentials for quarkonium have been calculated using lattice QCD in Refs.~\cite{Bal05,Bul19}. The $\Sigma_g^+$ potentials with string breaking of Ref.~\cite{Bal05} have been used in Refs.~\cite{Bic20,Bic21} to calculate the masses of bottomonium states and their widths into pairs of $S$-wave heavy-light mesons under the assumption of exact heavy-quark spin symmetry (HQSS). We denote pairs of $S$-wave heavy-light mesons by $B^{(\ast)}\bar{B}^{(\ast)}$, where $B^{(\ast)}$ is the pseudoscalar meson $B$ or the vector meson $B^\ast$. The effects of $B^{(\ast)}\bar{B}^{(\ast)}$ thresholds on the masses and widths of $Q\bar{Q}$ mesons can be calculated in a B\nobreakdash-O effective field theory \cite{Tar22}. In Ref.~\cite{Bru23}, it was shown that the transition rate between quarkonium and $B^{(\ast)}\bar{B}^{(\ast)}$ B\nobreakdash-O configurations can be expressed as the product of a string-breaking rate accessible by lattice QCD calculations and a multiplicative coefficient determined by the B\nobreakdash-O symmetries, that is, cylindrical symmetry, $CP$, and HQSS.

The decay of a $Q\bar{Q}$ meson into $B^{(\ast)}\bar{B}^{(\ast)}$ can proceed through the breaking of the flux tube connecting the $Q$ and $\bar{Q}$ by the creation of a $q\bar{q}$ pair, resulting in the formation of two heavy-light mesons $Q\bar{q}$ and $q\bar{Q}$. If lattice QCD calculations of hybrid B\nobreakdash-O potentials with string breaking were available, they could be used to calculate the $B^{(\ast)}\bar{B}^{(\ast)}$ decay widths of quarkonium hybrids. There are as yet no such lattice QCD calculations. However, one can apply the techniques of Ref.~\cite{Bru23} to the B\nobreakdash-O channels for quarkonium hybrids to extract selection rules for their decays into $B^{(\ast)}\bar{B}^{(\ast)}$ fixed solely by the B\nobreakdash-O symmetries.

Let us begin by considering a static $Q\bar{Q}$ pair with separation $\bm{r}$ in the presence of light QCD fields with general B\nobreakdash-O quantum numbers $\Lambda_\eta^\epsilon$. Let $\bm{J}_\textup{light}$ be the light QCD angular momentum and $J_\textup{light}$ its corresponding quantum number. In the case of zero $Q\bar{Q}$ separation, $\bm{r}=0$, the light QCD states can be grouped in multiplets with definite values of $J_\textup{light}$, $P_\textup{light}$, and $C_\textup{light}$, which are the quantum numbers corresponding to the light $O(3) \times C$ symmetry. Although these quantum numbers do not apply to the system of two static quarks with nonzero separation, they will become useful to determine the symmetries of the system when the motion of the heavy quarks is taken into account.

The value of $J_\textup{light}$ can be inferred by looking at the degeneracy of the static energy levels in the limit $\bm{r}\to 0$. The quarkonium B\nobreakdash-O potential $\Sigma_g^+$ is nondegenerate and has $J_\textup{light}=0$. The two lowest hybrid B\nobreakdash-O potentials, $\Pi_u$ and $\Sigma_u^-$ ($\bm{J}_\textup{light}\cdot\bm{\hat{r}}=\pm1$ and $0$), are degenerate for $\bm{r}\to0$ and correspond to $J_\textup{light}=1$. Parity and charge-conjugation are not, in general, symmetries of the light QCD fields in presence of two static sources. Only their combination $CP$ is. However, one can define quantum numbers $P_\textup{light}$ and $C_\textup{light}$ corresponding to these symmetries in the case $\bm{r}=0$. The value of $P_\textup{light}$ can be determined from the reflection quantum number $\epsilon$ of the $\Lambda=0$ member of the light QCD multiplet:
\begin{subequations}
\label{plightclighteq}
\begin{equation}
P_\textup{light}=\epsilon(-1)^{J_\textup{light}}.
\end{equation}
Then, since $(CP)_\textup{light}=+$ or $-$ corresponds to $\eta=g$ or $u$, respectively, $C_\textup{light}$ is easily determined as
\begin{equation}
C_\textup{light} = \eta P_\textup{light} = \eta \epsilon(-1)^{J_\textup{light}}.
\end{equation}
\end{subequations}
Following this prescription, it is shown that the hybrid B\nobreakdash-O potentials $\Pi_u$ and $\Sigma_u^-$ with $\epsilon=-$ and $\eta=u$ belong to the $1^{+-}$ representation of the light QCD symmetry group at $\bm{r}=0$.

From these B\nobreakdash-O quantum numbers, static B\nobreakdash-O quantum numbers that include the total $Q\bar{Q}$ spin $\bm{S}_{Q\bar{Q}}$ are automatically determined. We define the static angular momentum
\begin{equation}
\bm{J}_\textup{static} = \bm{J}_\textup{light} + \bm{S}_{Q\bar{Q}}
\label{stateq}
\end{equation}
whose quantum number is $J_\textup{static}$. The $CP$ quantum number of the $Q\bar{Q}$ meson is
\begin{equation}
CP = \eta (-1)^{S_{Q\bar{Q}} + 1}.
\label{cpeq}
\end{equation}
Specifically, for quarkonium hybrids with $\eta=u$ one has $CP=(-1)^{S_{Q\bar{Q}}}$. $CP=+$ corresponds to $S_{Q\bar{Q}}=0$, and the quarkonium hybrid B\nobreakdash-O configurations have $J_\textup{static}=1$. $CP=-$ corresponds to $S_{Q\bar{Q}}=1$, and the quarkonium hybrid B\nobreakdash-O configurations have $J_\textup{static}=0,1,2$.

One can define the spin $J$ of the $Q\bar{Q}$ meson by introducing the motion of the heavy quarks. The total angular momentum of the $Q\bar{Q}$ system in its rest frame is 
\begin{equation}
\bm{J} = \bm{L} + \bm{J}_\textup{static},
\end{equation}
where $\bm{L}$ is the relative orbital angular momentum of the $Q\bar{Q}$ pair and $\bm{J}_\textup{static}$ is the static angular momentum defined in Eq.~\eqref{stateq}. We denote the corresponding quantum numbers by $J$, $L$, and $J_\textup{static}$. The parity and charge-conjugation quantum numbers of the heavy quark-antiquark system are determined by the transformation of both the $Q\bar{Q}$ pair and the light QCD fields under these symmetries. Continuity at $\bm{r}=0$ implies that the light QCD state corresponding to a light $\Lambda_\eta^\epsilon$ B\nobreakdash-O channel, $\ket{\zeta_{\Lambda_\eta^\epsilon}(\bm{r})}$, transforms as
\begin{subequations}
\begin{align}
P\ket{\zeta_{\Lambda_\eta^\epsilon}(\bm{r})} &= P_\textup{light} \ket{\zeta_{\Lambda_\eta^\epsilon}(-\bm{r})}, \\
C\ket{\zeta_{\Lambda_\eta^\epsilon}(\bm{r})} &= C_\textup{light} \ket{\zeta_{\Lambda_\eta^\epsilon}(-\bm{r})},
\end{align}
\end{subequations}
for any $Q\bar{Q}$ separation $\bm{r}$, with $P_\textup{light}$ and $C_\textup{light}$ the quantum numbers of the light QCD symmetries at $\bm{r}=0$ defined in Eqs.~\eqref{plightclighteq}. The parity and charge-conjugation quantum numbers of the heavy quark-antiquark system are then
\begin{subequations}
\begin{align}
P &= P_\textup{light} (-1)^{L+1}, \\
C &= C_\textup{light} (-1)^{L + S_{Q\bar{Q}}}.
\end{align}
\end{subequations}
Their product agrees with Eq.~\eqref{cpeq}. Specifically, for quarkonium hybrids with $P_\textup{light}=+$ and $C_\textup{light}=-$ one has $P=(-1)^{L+1}$ and $C=(-1)^{L+S_{Q\bar{Q}}+1}$.

In general, quarkonium hybrid B\nobreakdash-O configurations could decay through string breaking into $B^{(\ast)}\bar{B}^{(\ast)}$ configurations with compatible light B\nobreakdash-O quantum numbers $\Pi_u$ and $\Sigma_u^-$. As indicated in Table~I of Ref.~\cite{Tar22}, the $B^{(\ast)}\bar{B}^{(\ast)}$ configurations are compatible with light B\nobreakdash-O quantum numbers $\Sigma_g^+$, $\Pi_g$, and $\Sigma_u^-$, but not $\Pi_u$. Combining this with the techniques pioneered in Ref.~\cite{Bru23} allows us to point out for the first time that quarkonium hybrids may decay into $B^{(\ast)}\bar{B}^{(\ast)}$ pairs, and that the decays proceed exclusively through string-breaking from the gluonic state with light B\nobreakdash-O quantum numbers $\Sigma_u^-$.

This observation produces a B\nobreakdash-O selection rule. In the limit of exact HQSS, quarkonium hybrid states form degenerate $J^{PC}$ multiplets. The first five multiplets are listed in Table~\ref{multable} (see Table~II of Ref.~\cite{Ber15}). Each of these multiplets corresponds to a bound state of either the $\Pi_u$ potential, the $\Sigma_u^-$ potential, or coupled $\Pi_u$ and $\Sigma_u^-$ potentials. The B\nobreakdash-O selection rule is that only hybrids in multiplets associated to either the $\Sigma_u^-$ potential or coupled $\Pi_u$ and $\Sigma_u^-$ potentials can decay into $B^{(\ast)}\bar{B}^{(\ast)}$. Hybrids in the $H_2$ and $H_5$ multiplets are forbidden to decay into $B^{(\ast)}\bar{B}^{(\ast)}$ since they are bound states of the $\Pi_u$ potential. On the other hand, these decays are allowed for hybrids in the $H_1$, $H_3$, and $H_4$ multiplets.

\begin{table}
\caption{\label{multable}Lowest five hybrid multiplets and their potential(s).}
\begin{ruledtabular}
\begin{tabular}{ccc}
Multiplet		& $J^{PC}$					&  Potential(s)				\\
\hline
$H_1$			& $1^{--}$, $(0,1,2)^{-+}$		& coupled $\Pi_u$ and $\Sigma_u^-$ 	\\
$H_2$			& $1^{++}$, $(0,1,2)^{+-}$		& $\Pi_u$ 					\\
$H_3$			& $0^{++}$, $1^{+-}$			& $\Sigma_u^-$ 				\\
$H_4$			& $2^{++}$, $(1,2,3)^{+-}$		& coupled $\Pi_u$ and $\Sigma_u^-$ 	\\
$H_5$			& $2^{--}$, $(1,2,3)^{-+}$		& $\Pi_u$						\\
\end{tabular}
\end{ruledtabular}
\end{table}

Since the B\nobreakdash-O symmetries allow $B^{(\ast)}\bar{B}^{(\ast)}$ decays for many hybrid $J^{PC}$ quantum numbers, the only remaining possibility for a suppression of all quarkonium hybrid decays into two $S$-wave heavy-light mesons would be a dynamical QCD suppression of the string-breaking rate for light B\nobreakdash-O quantum numbers $\Sigma_u^-$. However, the lattice QCD results in Ref.~\cite{Shi23} indicate that such dynamical suppression does not occur for the lowest $1^{-+}$ charmonium hybrid, which is a state in the $H_1$ multiplet. We therefore conclude that the decays of most quarkonium hybrids into two $S$-wave heavy-light mesons are not suppressed, in contrast to conventional wisdom from hybrid meson decay models.

In Ref.~\cite{Bru23}, it is argued that a formal B\nobreakdash-O approximation must include the effects of the $B^\ast$-$B$ mass splitting in addition to the kinetic energies of $Q$ and $\bar{Q}$. The inclusion of the $B^\ast$-$B$ mass splitting breaks HQSS, so that the degeneracy of the hybrid multiplets is broken. It is therefore useful to derive a second B\nobreakdash-O selection rule that allows HQSS-suppressed $B^{(\ast)}\bar{B}^{(\ast)}$ decays of some states in the $H_2$ and $H_5$ hybrid multiplets.

One can use the Fierz identities to show that a configuration with light B\nobreakdash-O quantum numbers $\Sigma_u^-$ created from a $B^{(\ast)}\bar{B}^{(\ast)}$ source corresponds to $J_\textup{light}=0$. Thus, it follows immediately from angular momentum that a quarkonium hybrid B\nobreakdash-O configuration with total $Q\bar{Q}$ spin $S_{Q\bar{Q}}$ can decay only into a $B^{(\ast)}\bar{B}^{(\ast)}$ B\nobreakdash-O configuration whose static angular momentum satisfies $J_\textup{static}=S_{Q\bar{Q}}$. As the static angular momentum quantum number $J_\textup{static}$ of a $B^{(\ast)}\bar{B}^{(\ast)}$ B\nobreakdash-O configuration coincides with the total spin quantum number $S$ of the di-meson pair, one has the B\nobreakdash-O spin selection rule
\begin{equation}
S = S_{Q\bar{Q}},
\label{selruleq}
\end{equation}
where $S$ is the total spin of the two mesons in the $B^{(\ast)}\bar{B}^{(\ast)}$ decay channel and $S_{Q\bar{Q}}$ is the total $Q\bar{Q}$ spin of the decaying hybrid meson. Therefore, quarkonium hybrids are forbidden to decay into $B^{(\ast)}\bar{B}^{(\ast)}$ if the di-meson channels available under the B\nobreakdash-O spin selection rule in Eq.~\eqref{selruleq} cannot generate the corresponding $J^{PC}$. Thus, in the $H_2$ multiplet, decays into pairs of $S$-wave heavy-light mesons are forbidden for $1^{++}$ and $0^{+-}$ but suppressed by HQSS for $1^{+-}$ and $2^{+-}$. In the $H_5$ multiplet, decays into pairs of $S$-wave heavy-light mesons are forbidden for $2^{--}$ but suppressed by HQSS for $(1,2,3)^{-+}$.

Translating the model-independent B\nobreakdash-O description of $Q\bar{Q}$ meson decays into the language of decay models, we can say that the B\nobreakdash-O picture is actually consistent with Page's selection rule in Eq.~\eqref{pageq}. In fact, the B\nobreakdash-O symmetries imply $S_{q\bar{q}}=1$ for quarkonium decays (B\nobreakdash-O channel $\Sigma_g^+$) and $S_{q\bar{q}}=0$ for quarkonium hybrid decays (B\nobreakdash-O channel $\Sigma_u^-$). In most hybrid meson decay models, the assumption $S_{q\bar{q}}=1$ is motivated by a physical picture in which the gluonic excitation decays into a light $q\bar{q}$ pair through a spin-conserving process while the heavy quarks act as spectators. However this does not take into account the fact that the $Q\bar{Q}$ pair, which is essentially stationary on the time scale for the breaking of the flux tube, acts as a static color dipole that breaks rotational symmetry for the light QCD fields. This allows deviations from the naive expectation that $S_{q\bar{q}}$ should coincide with the spin of the decaying gluonic excitation. In the presence of a static $Q\bar{Q}$ pair, rotational symmetry is broken to the B\nobreakdash-O symmetry group, which is the only correct symmetry group of a heavy quark-antiquark system. The assignment $S_{q\bar{q}}=0$, that allows decays into a pair of $S$-wave mesons, was also proposed in a recent quarkonium hybrid decay model inspired by the B\nobreakdash-O symmetries \cite{Bru19d}.

It is interesting to observe that quarkonium and quarkonium hybrid configurations may mix with each other through their common coupling to $B^{(\ast)}\bar{B}^{(\ast)}$, once the $B^\ast$-$B$ mass splitting is taken into account. The $B^{(\ast)}\bar{B}^{(\ast)}$ operators have the schematic form $(\bar{Q}\Gamma q)(\bar{q}\Gamma^\prime Q)$, where $\Gamma$ and $\Gamma^\prime$ are $4\times4$ Dirac matrices. A Fierz transformation can be used to expand them into products of operators with the schematic form $(\bar{Q}\Gamma Q)(\bar{q}\Gamma^\prime q)$. For quarkonium, the $\Sigma_g^+$ light B\nobreakdash-O state mixes with the $\bar{q}P_- \bm{\gamma}\cdot\bm{\hat{r}} q$ component of the $B^{(\ast)}\bar{B}^{(\ast)}$ operators. For quarkonium hybrids, the $\Sigma_u^-$ light B\nobreakdash-O state mixes with the $\bar{q}P_-\gamma_5 q$ component of the $B^{(\ast)}\bar{B}^{(\ast)}$ operators. Breaking of HQSS through the $B^\ast$-$B$ mass splitting mixes the $\bar{q}P_- \bm{\gamma}\cdot\bm{\hat{r}} q$ and $\bar{q}P_-\gamma_5q$ operators with each other; see Ref.~\cite{Bru23}. Hence, the combination of HQSS breaking and string breaking implies some degree of mixing between quarkonium and quarkonium hybrids. This mixing should be expected to be quite small, since it proceeds through two string-breaking transitions and suffers a suppression factor from HQSS breaking.

The effects of the $B^{(\ast)}\bar{B}^{(\ast)}$ coupling on the quarkonium and quarkonium hybrid spectra can be studied in B\nobreakdash-O with coupled channels. Specifically, in the so-called diabatic representation of B\nobreakdash-O, the B\nobreakdash-O potentials with string breaking are translated into a multichannel potential matrix \cite{Bru20}. For quarkonium, the string-breaking transition terms in the diabatic potential matrix can be determined as shown in Ref.~\cite{Bru23}. For quarkonium hybrids, the string-breaking transition terms can be determined analogously.

In the limit of exact HQSS, one can predict the branching ratios of a quarkonium hybrid into different pairs of $S$-wave mesons by taking the square of the ratios of the corresponding string-breaking transition terms, which is determined solely by the B\nobreakdash-O symmetries. We predict a value of $1$ for the branching ratio of a $1^{-+}$ quarkonium hybrid into $B^\ast\bar{B}$ and $B^\ast\bar{B}^\ast$, which is compatible within the large uncertainty for the ratio $0.6 \pm 0.5$ of the corresponding decay widths of the lowest $1^{-+}$ charmonium hybrid calculated using lattice QCD in Ref.~\cite{Shi23}.

The diabatic potential matrix including quarkonium, quarkonium hybrid, and di-meson channels, as well as its numerical application, will be dealt with in future research. The explicit calculation of the hybrid B\nobreakdash-O potentials with string breaking using lattice QCD would be essential for quantitative predictions of the spectrum and decay widths. Lattice QCD calculations of charmonium hybrid mesons with dynamical charm quarks \cite{Man01,Liu12,Che16,Shi23}, although currently available only for unphysically large pion masses, can provide an essential input against which the predictions of the B\nobreakdash-O approximation can be tested.

The B\nobreakdash-O selection rules derived here may serve as a guide for future experimental searches of quarkonium hybrids. Although not as constraining as the selection rules from hybrid meson decay models, the B\nobreakdash-O selection rules provide a specific pattern of suppression of the decays of quarkonium hybrids into $B^{(\ast)}\bar{B}^{(\ast)}$ that may help identify them in experiments. Note that the $0^{+-}$ hybrid in the $H_2$ multiplet has manifestly exotic quantum numbers and its decays into $B^{(\ast)}\bar{B}^{(\ast)}$ are forbidden, so it should be relatively narrow. The identification of $X(3872)$ as the $1^{++}$ hybrid in the $H_2$ multiplet is disfavored by its mass, but it can be ruled out by its strong observed coupling to $D^\ast\bar{D}$. Finally, the B\nobreakdash-O selection rules imply that quarkonium hybrid states that decay into $B^\ast\bar{B}^\ast$ can additionally decay into either $B^\ast\bar{B}$ or $B\bar{B}$, but never both. Hence, the observation of a resonance in the $B\bar{B}$ and $B^\ast\bar{B}^\ast$ channels but not in the $B^\ast\bar{B}$ channel, or its observation in the $B^\ast\bar{B}$ and $B^\ast\bar{B}^\ast$ channels but not in the $B\bar{B}$ channel, may serve as a distinctive experimental signature of a quarkonium hybrid meson.

\acknowledgments{%
This work would not have been possible without constant support from E. Braaten. I acknowledge useful discussions with P. Gonz\'alez. This research was supported by the U.S. Department of Energy under Grant No.~\texttt{DE-SC0011726}.
}

\bibliography{diabhybridbib}

\end{document}